\documentclass{ws-procs9x6}

\begin{document}

\title{Using boundary methods to compute the Casimir energy}

\author{F.C. Lombardo$^1$, F.D.  Mazzitelli$^1$, and P.I. Villar$^{1,2}$}

\address{$^1$ Departamento de F\'\i sica {\it Juan Jos\'e
 Giambiagi}, FCEyN UBA, Facultad de Ciencias Exactas y Naturales,
 Ciudad Universitaria, Pabell\' on I, 1428 Buenos Aires, Argentina \\
$^2$ Computer Applications on Science and Engineering Department,
Barcelona Supercomputing Center (BSC), 29, Jordi Girona 08034
Barcelona, Spain}

\begin{abstract}
We discuss new approaches to compute numerically the Casimir interaction
energy for waveguides of arbitrary section, based on the boundary methods
traditionally used to compute eigenvalues of the 2D Helmholtz equation.
These methods are combined with the Cauchy's  theorem in order to perform
the sum over modes. As an illustration, we describe a point-matching technique
to compute the vacuum energy for waveguides containing media with different permittivities. 
We present
explicit numerical evaluations for perfect conducting surfaces
in the case of concentric corrugated cylinders and a circular cylinder inside an
elliptic one.
\end{abstract}


\bodymatter


\section{Introduction}

In this paper we will be concerned with the numerical calculation of
the Casimir interaction energy in geometries with translational
invariance along one direction, i.e. very long cylinders of arbitrary
section.
For the sake of simplicity, we will first discuss  the case of a
massless quantum scalar field that satisfies Dirichlet or Neumann
boundary conditions on the surfaces of the cylinders. As we will see,
in some particular situations the generalization to the
electromagnetic field and/or more general boundary conditions
will be straightforward.

The Casimir energy is formally given by
\begin{equation}
 E_{12}(\sigma)= \lim_{\sigma\to 0}{1\over 2}\sum_p(e^{-\sigma
w_p} w_p-e^{-\sigma \tilde w_p} \tilde w_p)\, ,
 \label{modes}
\end{equation}
where $w_p$ are the eigenfrequencies of the scalar field
satisfying the appropriate boundary conditions on the surfaces of
the shells, and $\tilde w_p$ are those corresponding to a situation
in which the distances between the shells is very large.
The subindex $p$ denotes the set of quantum
numbers associated to each eigenfrequency. We have
introduced an exponential  cutoff for
high frequency modes.

For the particular geometry considered here, $p=(n,k_z)$ and the
eigenfrequencies are
of the form $\omega_{n,k_z}=\sqrt{k^2+\lambda_n^2}$,
where $k_z$ is a continuous variable associated to the translational invariance
along the z-direction and $\lambda^2$ are the eigenvalues of the
Laplacian on the two-dimensional transversal section $\Sigma$
contained in the plane (x,y):
\begin{equation}
\Delta^2 u=-\lambda^2 u\,\, .
\label{Helmholtz}
\end{equation}
The eigenfunctions $u({\bold x})$ satisfy Dirichlet or Neumann boundary conditions
on $\Gamma$, the boundary of $\Sigma$.

The Helmholtz equation (\ref{Helmholtz}) arises in many branches of
physics, from the vibration of membranes to quantum billiards, and
there are a plethora of methods to compute numerically its
eigenfunctions and eigenvalues \cite{Kuttler}. Among them, the "boundary methods" are 
based on the following strategy: the solution $u$ is written as a (finite) linear combination of
basis functions that
satisfy Helmholtz equation inside $\Sigma$. The coefficients of the linear
combination are chosen in such a way that the boundary conditions are
satisfied at a
finite number of points on $\Gamma$. The linear system of equations that determine the
coefficients has a non trivial solution only for some particular values of $\lambda$, 
the eigenvalues of the system.

For example, in the Point Matching
Method (PMM)\cite{pmm}, one expands the eigenfunction $u$ in  terms of a basis
of solutions of the Helmholtz equation in free space
$\varphi_j^{(\lambda)}({\bold x})$
\begin{equation}
u({\bold x})=\sum_{j=1}^\infty a_j\varphi_j^{(\lambda)}({\bold x})\, .
\end{equation}
In the numerical calculation this expansion is truncated at  given
$j=N$, and the
boundary conditions are imposed on $N$ points on $\Gamma$. These
boundary conditions become a set of homogeneous, linear equations for
the unknown coefficients $a_j$ ($M a=0$, with $M$ a
$\lambda$-dependent $N\times N$ matrix)  which has  nontrivial
solutions  only when $det M=0$. The last equation can be used to
determine numerically the eigenvalues  $\lambda_n$.

In a  similar approach, known as the Method of Fundamental Solutions
(MFS) \cite{mfs},
the eigenfunction $u$ is expanded in terms of solutions of the
Helmholtz equation
with a point source at an arbitrary location ${\bold s_j}$, that we
denote by $u_\lambda({\bold x},{\bold s_j})$
\begin{equation}
u({\bold x})=\sum_{j=1}^\infty b_j u_\lambda({\bold x}, {\bold s_j}) \, .
\end{equation}
If the sources are located {\it outside} $\Sigma$, this is a
solution of the homogeneous
Helmholtz equation {\it inside} $\Sigma$. Once again, the sum is
truncated at $j=N$, and the coefficients $b_j$ are determined by
solving the linear system that results when imposing the boundary
conditions on  a
finite number of points on $\Gamma$. The roots of the determinant of
the associated matrix
are the eigenfrequencies of the problem. This is the simplest version of
the MFS, in which the locations of the sources are fixed.

One can find in the literature discussions about spurious solutions,
improvements and
alternative methods  to find the eigenvalues. We refer the reader to
Refs. \cite{Kuttler,others}
for more details.

A crucial point is that, at a practical level, the knowledge of the spectrum of
the Helmholtz equation is not enough
to compute the Casimir energy. The reason is that the numerical
evaluation of the  sum over modes in Eq.(\ref{modes}) is extremely
unstable \cite{Johnson}, and one has to subtract very large numbers to
compute the
finite interaction energy. The calculation is complicated even  for
the simplest case of  Casimir effect in $1+1$
dimensions.

Instead of performing explicitly the summation, it is far more
efficient to combine the methods mentioned previously with the
Cauchy's theorem
\begin{equation}
{1\over 2\pi i} \int_{C} \,dz \;  z \; e^{-\sigma
z} {d\over dz} \ln f(z)=\sum_i z_i \;
e^{-\sigma z_i} \; ,
\end{equation}
where $f(z)$ is an analytic function in the complex $z$
plane within the closed contour ${C}$, with simple zeros at
$z_1, z_2, \dots$ within ${C}$. We use this result to
replace the sum over the eigenvalues of the Helmholtz equation
in the Casimir energy Eq.(\ref{modes}) by a contour integral.
In this way, it is not necessary to solve numerically the equation $det M=0$ for
the eigenvalues, but to take  $f=det M$ in the Cauchy's theorem.  In
other words,
if in the numerical method to solve the Helmholtz equation the
eigenvalues are the roots of a
given function, one can integrate this function in the complex plane
in order to get the Casimir energy. The combination of the use of
numerical methods
to compute the eigenvalues with the Cauchy's theorem is the main idea
we want to put
forward in this paper.

In the next Section we will describe the simplest version of the PMM
to a situation
in which the surfaces separate regions of different permittivities, generalizing
our previous results \cite{PRDpm} for perfect conductors. In Sections
3 and 4  we will review
some numerical evaluations of the Casimir interaction energy for
perfect conductors. In Section 5
we include our final remarks.


\section{Point-Matching Numerical Approach}
\label{method}

A media-separated waveguide presents an interesting setup for the
application of the PMM. This technique has been
widely used to solve eigenvalue problems in many areas of
engineering science \cite{pmm}. The boundary conditions are imposed
at a finite number of points around the periphery of both media.

\begin{figure}[ht]
\centering
\includegraphics[width=6.5cm]{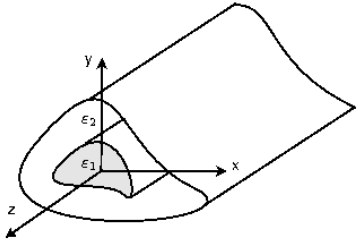}
\caption{A two-separated media waveguide in which one conductor
encloses two different dielectric media. Each has arbitrary cross
section.} \label{fig1}
\end{figure}

For the sake of concreteness, we will bear in mind the situation in which
one perfect conductor encloses two dielectric media, as shown in
Fig.\ref{fig1}, although the method could be applied to more general
cases.

The general solution of the Helmholtz equation
in region I  (inside the inner cylinder) is
\begin{equation}
u=\sum_{m} A_m J_m(\lambda^{(I)} r) e^{i m \theta}, \label{c1}
\end{equation}
while in region II (annular region)
\begin{equation}
u =\sum_{m} [B_m J_m(\lambda^{(II)} r) + C_m H^{(1)}_m(\lambda^{(II)} r)]
e^{i m \theta},\label{c2}
\end{equation}
where $(r,\theta)$ are polar coordinates, and $J_m$ and
$H^{(1)}_m$ are the $m$-th order Bessel functions. The constants
$A_m, B_m$ and $C_m$  are  determined by the boundary conditions.
In both equations we have defined
$\lambda^{(a)}= \sqrt{\epsilon_a \omega^2 - k_z^2}$.

We assume the outer surface to be a "perfect conductor", and impose
Dirichlet boundary conditions on a finite number of points $(r_q,\theta_q)$ of
${C}_2$:
\begin{equation}
0= \sum_{m=-S}^{S} [B_m J_m(\lambda^{(II)}  r_q) + C_m H_m^{(1)}(\lambda^{(II)}
r_q)] e^{i m \theta_q}, \label{perfectcond}\end{equation}
(alternatively, for the TE modes of the electromagnetic field,
one should impose
Neumann boundary conditions).
The surface $C_1$ as a
dielectric interphase separating media $\epsilon_1$ and
$\epsilon_2$, and therefore we impose continuity of the field
and its derivative:
\begin{eqnarray}
\sum_{m = -S}^{S} A_m J_m(\lambda^{(I)}  r_p) e^{i m \theta_p} &=&
\sum_{m = - S}^S [B_m J_m(\lambda^{(II)}  r_p) + C_m H^{(1)}_m(\lambda^{(II)}
r_p)] e^{i m \theta_p}  \\
\sum_{m = -S}^{S} A_m J_m'(\lambda^{(I)}  r_p) e^{i m \theta_p} &=&
\frac{\lambda^{(II)} }{\lambda^{(I)} }\sum_{m = - S}^S [B_m
J_m'(\lambda^{(II)}  r_p)
+ C_m H'^{(1)}_m(\lambda^{(II)}  r_p)] e^{i m \theta_p},\nonumber
\label{continuity}\end{eqnarray}
where $(r_p,\theta_p)$
are points on the curve ${C}_1$.

The boundary conditions can be written, in matrix form,
as
\begin{eqnarray}
 0 &=& N_1 B + N_2 C  ,\nonumber \\
 R_1 A &=& M_1 B + M_2 C, \nonumber \\
R_2 A &=& M_1'B + M_2'C. \label{setmatriz}
\end{eqnarray}
Eliminating the coefficients $A_m$ we end with
\begin{equation}
 N_1 B + N_2 C = 0, \,\, \,\,
 P_1 B + P_2 C  = 0,
 \label{syst}
\end{equation}
where $P_1$ and
$P_2$ can be written as
\begin{equation}
P_1 = M_1 - R_1 R_2^{-1} M_1',  \,\,\,\,\,
P_2 = M_2 - R_1 R_2^{-1} M_2'\, .
\end{equation}

It is worthy to note that as $R_1 R_2^{-1}$ is
proportional to $\lambda_2/\lambda_1$, then  $R_1 R_2^{-1}\rightarrow 0$ when
$\epsilon_1 \rightarrow \infty$. Thus, the matrices $P_1 \rightarrow
M_1$ and $P_2 \rightarrow M_2$, re-obtaining in this way, the usual
perfect conductor wave-guide case studied in \cite{PRDpm}.

For the system of Eq.(\ref{syst}) to have non trivial solutions,  the
determinant must be zero, i.e.
\begin{equation}
 det\bigg[ \begin{array}{c c}
 N_1 & N_2  \\
P_1 & P_2
    \end{array}
\bigg] = det P_2 \cdot det N_1 \cdot  det (1 - N_2 P_2^{-1} P_1N_1^{-1})=0 ~.
\end{equation}
This equation determines the eigenfrequencies associated to the
geometry.  However, as already mentioned, in
order to compute the Casimir energy it is not necessary to find each
eigenvalue but to integrate the determinant
$Q=det (1 - N_2 P_2^{-1} P_1N_1^{-1})$
in the complex plane.

We have developed a numerical Fortran routine in order to evaluate
the Casimir interaction energy  in the case in which the field satisfies
Dirichlet or Neumann boundary conditions
on both curves $C_1$ and $C_2$.
In this case one should consider  the fields only in region II with
$\epsilon_2=1$.
After some straightforward steps one can re-write
the Casimir energy as a single integral in the imaginary axis
$iy=\lambda^{(II)}$.
For Dirichlet boundary conditions the result is
\begin{equation}
E_{12}={L\over 4\pi} \int_{0}^{\infty} dy \ y\ln Q(iy)  \; ,
\label{xxx}
\end{equation}
while for Neumann boundary conditions one can derive a similar
expression with a different function $Q$.
It is worth to stress that these Casimir energies correspond to those of
TM and TE modes of the electromagnetic field in the presence of
perfect conductors.

\section{Cylindrical rack and pinion}
\label{rack}

When two concentric cylinders have corrugations, the vacuum energy
produces a torque that could, in principle, make one cylinder rotate
with respect to the other. This ``cylindrical rack and pinion'' has
been  proposed in Ref. \cite{brasilia}, where the torque has been
computed using the proximity force approximation. It was further
analyzed in \cite{CaveroI},  where the authors obtained perturbative
results for Dirichlet boundary conditions in the limit of small
amplitude corrugations. In this Section, we numerically evaluate the
Casimir interaction energy for two concentric corrugated, perfect
conductor cylinders. The cylinders have radii $a$ and $b$, and we
will denote by $r_-=b-a$ the mean distance between them and by
$r_+=a+b$ the sum of the radii. We will use the notation $\alpha =
b/a$. The
points in the mesh, that give us the corrugated cylinder boundaries, 
are described by the following functions:
\begin{equation}
h_a(\theta) = h \sin(\nu \theta)~~~;~~~ h_b(\theta) = h \sin(\nu
\theta + \phi_0),
\end{equation}
where $h$ is the corrugation amplitude and $\nu$ is the frequency
associated with these corrugations. The Casimir torque can be
calculated by taking the derivative of the interaction energy with
respect to the shifted angle ${\cal T}= - \partial E_{12}/
\partial \phi_0$.

\begin{figure}[h!t]
\centering
\includegraphics[width=8.6cm]{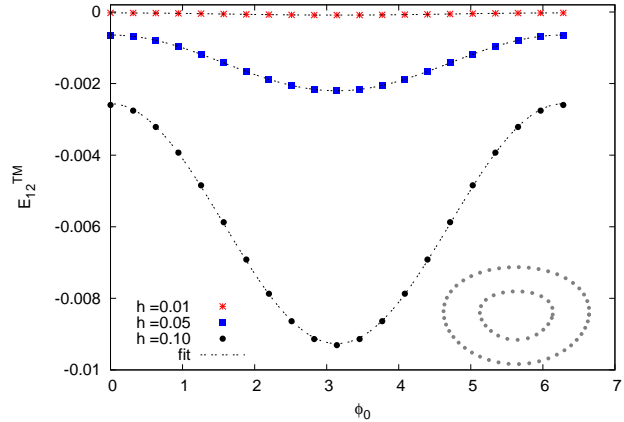}
\caption{Casimir interaction energy (TM modes) as a function of
$\phi_0$ for $\alpha =2$ and different values of the perturbation
$h$. The different shaped dots are the numerical data
obtained with our program while the lines represent the numerical
fit of each curve. Energies are measured in units of $L/a^2$, and 
distances in units of $a$.}
\label{fig10}
\end{figure}

In Fig.\ref{fig10} we show the numerical evaluation of the TM
Casimir interaction energy for this geometry. The plot shows the
results obtained using the  PMM with  $\alpha=2$
and corrugation frequency $\nu=3$, for different values of the
amplitude of the corrugation $h$.  As expected the amplitude of the
oscillations grows with $h$. For each value of $h$ we have performed
a numerical fit of the data in order to compare with the analytical
prediction. With dotted lines we have
plotted the fit $y(x)=A*cos(x)$ for each curve in Fig.\ref{fig10}. 
The agreement between dots and dotted lines is extremely good.
Similar results can be obtained for the Neumann (TE) modes (see 
\cite{PRDpm} for details).

\begin{figure}[h!t]
\centering
\includegraphics[width=8.6cm]{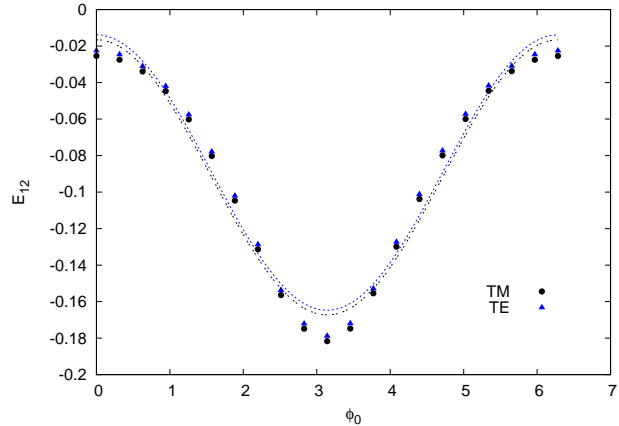}
\caption{Casimir interaction energy (TE and TM modes) as a function
of $\phi_0$ for $\alpha=2$, $\nu = 3$ and $\tilde h = 0.3$. The
different shaped dots are the numerical data obtained by our
program while the line represents the numerical fit of each curve.
In this case, the plot shows that the exact result cannot be fitted
by a function $y(x) = A * \cos(x)$. Energies are measured in units
of $L/a^2$.} \label{fig12}
\end{figure}

It is worth to remark that, when the amplitude of the corrugation is
not very small, the exact results cannot be reproduced with a simple
fit of the form $y(x)= A*\cos(x)$. This is illustrated in
Fig.\ref{fig12}, where we see that, for the biggest corrugated
amplitude $\tilde h = h/a = 0.3$, the exact result differs from the cosine function \cite{PRDpm}.

\section{Outer conductors with focal lines: cylinder inside an ellipse}
\label{elipses}

Some time ago, there was a conjeture \cite{ford} based on a
geometric optics approximation, about the possibility of focusing
vacuum fluctuations in parabolic mirrors. It was argued that a
parabolic mirror is capable of focusing the vacuum modes of the
quantized electromagnetic field,  therefore creating large physical
effects near the mirror's focus. With this motivation, in this Section we shall evaluate the Casimir
interaction energy for configurations in which the outer conducting
shell has a cross section that contains focal points.

We will consider one small inner cylinder and an outer ellipse. We will 
denote by $a$ the radius of the
inner cylinder, by $b_1$ and $b_2$ the minor and major semiaxes of
the ellipse, respectively, and by $f$ the distance between the foci
and the center of the ellipse. The coordinates of the center of the
cylinder with respect to the center of the ellipse will be
$(\epsilon_x,\epsilon_y)$. We will use an additional tilde to denote
adimensional quantities, i.e distances in units of $a$: $\tilde
b_i=b_i/a\, , \tilde f=f/a$, etc.

\begin{figure}[h!t]
\centering
\includegraphics[width=8.6cm]{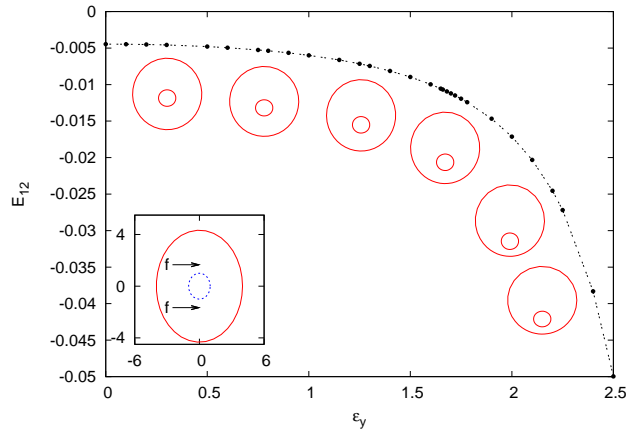}
\caption{Numerical evaluation of the Casimir interaction energy for
an inner cylinder an eccentric outer ellipse, as a function of the
position of the cylinder along the vertical axis. Energies are
measured in units of $L/a^2$.} \label{fig15}
\end{figure}

For this configuration, we use a mesh where with an inner cylinder, and an 
outer ellipse with semiaxes $\tilde b_1 = 4$ and $\tilde b_2 = 4.33$. The ellipse
has two focal points at  $\tilde f= 1.66$. We present the results
for the Casimir energy in Fig.\ref{fig15}.

From Fig.\ref{fig15} it is possible to see that there is an unstable
equilibrium position at the origin under displacements of the inner
cylinder along the (vertical)  $\epsilon_y$ direction. As expected,
it is also possible to check that the energy grows as well as the
cylinder gets closer to the surface of the outer ellipse.
Fig.\ref{fig15} also shows a monotonic behaviour of the energy as a
function of the position, even when passing through the focus.
So we do not see a focusing of vacuum fluctuations near the focus
of the ellipse.
However, in order to confirm this result  one should consider much 
smaller inner cylinders, in order to explore shorter wavelengths. This 
will require much more computational effort. 

Finally, we have also checked \cite{PRDpm} that there is an unstable equilibrium position
at the origin when moving the inner cylinder in the (horizontal)
$\epsilon_x$ direction.

\section{Final remarks}
\label{conc}

We have  presented new numerical methods to compute the
vacuum energy for
arbitrary geometries with translational invariance. The approach is
based on the use of traditional boundary methods to compute
eigenvalues of the two dimensional Helmholtz equation, combined with Cauchy's 
theorem.

As a particular example, we have  described a straightforward  version of the 
point-matching method to compute the Casimir interaction energy for 
a waveguide with different permittivities, and reviewed some numerical calculations
for perfect conductors. In all examples,  for the numerical calculations
we have chosen
pair of  points with the
same angular coordinate with respect to the inner cylinder. For less
symmetric configurations,
and when the surfaces of both conductors are closer to each other, it
will be necessary to
consider grids with a larger number of points, and to  optimize their
positions.
As in the applications to acoustic or classical electromagnetism,
special care must be taken for surfaces with pronounced edges, clefts
or "handles", where the point-matching technique may not be accurate
to determine the eigenfrequencies. In these cases, more sophisticated 
approaches \cite{others} could be necessary to optimize the 
numerical evaluation and to avoid spurious solutions.

We would like to thank Kim Milton for the
organization and his kind hospitality during QFEXT09. This work has
been supported by CONICET, UBA and ANPCyT, Argentina.

\end{document}